# Abnormally low thermal conductivity of 2D selenene: An *ab initio* study

Gang Liu[1*], Zhibin Gao[2*], Guo-Ling Li[1], Hui Wang[1]

[1]School of Physics and Engineering, Henan University of Science and Technology, Luoyang 471023, People's Republic of China

[2]Department of Physics, National University of Singapore, Singapore 117551, Republic of Singapore

## Abstract

The lattice thermal conductivity and thermal transport properties of 2D α-selenene are investigated based on the first-principles calculations. The isotropic in-plane thermal conductivity is as low as 3.04 W m$^{-1}$ K$^{-1}$ at room temperature, even abnormally lower than α-tellurene which processes analogous configuration and lower Debye temperature. We find this abnormal phenomenon reasonably stems from the larger anharmonicity of acoustic phonon branch. Moreover, the phonon spectra, elastic properties, and related thermal properties are also exhibited. Acoustic phonons contribute mainly to the total thermal conductivity. Furthermore, size effect, boundary effect, the total phase space for three-phonon processes, phonon group velocity and relaxation time are further investigated, and the last one is unveiled to be the key ingredient of thermal transport in 2D selenene.

## Introduction

Since the successful isolation of graphene in 2004,[1-3] Two-dimensional (2D) materials have been intensively investigated in recent years. Their intriguing properties and the potential applications result in the fast expansion of the family. Beyond graphene, numerous elemental 2D materials have been predicted and synthesized, such as borophene, group IV and group V elemental 2D materials.[4-11] However, there were few investigations about group VI elemental 2D materials reported,[12] until the 2D tellurene monolayer has been predicted and synthesized successfully on highly oriented pyrolytic graphite (HOPG) substrates by using molecular beam epitaxy.[13,14] A great many investigations about group VI elemental 2D materials have been reported especially for tellurene, and the stable structures of monolayer Se (named selenene) are also predicted theoretically.[12,13,15-18] For instance, it is found that square selenene has low thermal conductivity[15] and interesting electronic structures with two gapped semi-Dirac cones in Brillouin zone, displaying non-trivial topological properties.[12] Large-size 2D Se nanosheet is also synthesized in experiment, while the Se nanosheet phototransistors show an excellent photo responsivity, showing great potential in electronic and optoelectronic applications.[19]

*Corresponding author: Gang Liu, Email: liugang8105@gmail.com

*Corresponding author: Zhibin Gao, Email: zhibin.gao@nus.edu.sg

Thermal conductivity and transport property are important in the practical applications for materials. Furthermore, thermoelectric property is a hot research focusing of 2D materials, since a great number of 2D materials show high thermoelectric performance.[17-23] It is well known that low thermal conductivity is very important for thermoelectric materials, since the figure of merit $zT$, which is used to measure the thermoelectric efficiency, is inversely proportional to thermal conductivity. Specifically, $zT$ of a thermoelectric material is expressed as $zT = S^2\sigma T/(\kappa_e + \kappa_L)$, where $S$, $\sigma$, $T$, $\kappa_e$ and $\kappa_L$ are the Seebeck coefficient, electric conductivity, absolute temperature, electronic thermal conductivity and lattice thermal conductivity, respectively. Furthermore, usually $\kappa_L$ is dominant and $\kappa_e$ is comparatively small, which can be ignored safely for semiconductors. Based on the expression above, it is concluded that high thermoelectric performance needs not only a high power factor ($S^2\sigma$) but also a low thermal conductivity as well, especially a low value of $\kappa_L$. For instance, theoretical investigation shows α-tellurene processes high $zT$ as the low $\kappa_L$.[17] Thus, the investigation of $\kappa_L$ is an urgent need for the potential application of α-selenene as a thermoelectric material.

In this work, we focus on the $\kappa_L$ and thermal transport properties of α-selenene. The lattice thermal conductivity of α-selenene is investigated by using first-principles calculations theoretically, based on Boltzmann transport equation (BTE). It is found α-selenene possesses a quite low $\kappa_L$ of 3.04 W m$^{-1}$ K$^{-1}$, lower than that of α-tellurene,[17] which has similar structure and lower Debye temperature. This abnormal phenomenon reasonably stems from the stronger anharmonicity of acoustic phonons compared with the α-tellurene, as shown by Grüneisen parameters. Its phonon spectra, mechanical properties, Debye temperature are also exhibited. The size effect and boundary effect are studied by cumulative and size-dependent $\kappa_L$ respectively. Furthermore, the total phase space for three-phonon processes, the relaxation times and group velocities of phonon are investigated to explore the behind mechanism of low $\kappa_L$. It is concluded reasonably that relaxation time plays the dominant role in the thermal transport for 2D α-selenene.

## Computational and Theoretical Methods

The structure, electronic structure, and energy are calculated by using the Vienna ab initio simulation package (VASP),[24-26] based on density functional theory (DFT). The local density approximation (LDA) is chosen for exchange-correlation functional.[27] A plane-wave basis set is employed with kinetic energy cutoff of 500 eV. A Monkhorst-Pack[28] k-mesh of 10 × 10 × 1 is used to sample the Brillouin zone (BZ) during the optimization. All geometries are fully optimized with the energy and the force convergence criterions of 10$^{-8}$ eV and 10$^{-4}$ eV Å$^{-1}$. The vacuum space of at least 20 Å is kept along the $z$ direction, which is enough to avoid the interactions between periodical images.

The harmonic and anharmonic interatomic force constants are obtained by Phonopy[29] and ShengBTE.[30] A 5 × 5 × 1 supercell with a 2 × 2 × 1 k-mesh for sampling is used to obtain the harmonic interatomic force constants. For the anharmonic interatomic force constants, the same size supercell is adopted while the interactions are taken into account up to the tenth nearest neighbors. After careful test, we chose a 111 × 111 × 1 q-mesh to ensure the convergence of thermal conductivity.

Within the Boltzmann transport equation (BTE), the in-plane $\kappa_L$ of isotropic material can be calculated as a sum of contribution of all the phonons with mode $\lambda$ and wave vector **q**, expressed as:

$$\kappa_L = \frac{1}{V} \sum_{\lambda,\mathbf{q}} C_{\lambda,\mathbf{q}} (v_{\lambda,\mathbf{q}}^\alpha)^2 \tau_{\lambda,\mathbf{q}}^\alpha, \qquad (1)$$

where $V$ is the volume of the cell. For the phonon with mode $\lambda$ and wave vector **q**, $C_{\lambda\mathbf{q}}$ is the heat capacity, $v_{\lambda,\mathbf{q}}^\alpha$ and $\tau_{\lambda,\mathbf{q}}^\alpha$ are the group velocity and relaxation time along the $\alpha$ direction, respectively. Eq. (1) can be well solved using the ShengBTE code with iterative scheme,[30] and a great number of previous investigations show its reliability and validity.[17,31-35]

## Results and discussions

The optimized structure of α-selenene is shown in Fig. 1(a), with the lattice constant $a$ and buckling height $h$ of 3.650 and 3.113 Å, which are very closed to previous results.[16] It possesses a 1T-MoS$_2$-like structure belonging to $P\bar{3}M1$ (164) symmetry group, showing isotropic pattern in the 2D plane. Similar to α-tellurene, the outer and centered Se atoms have the coordination numbers of 3 and 6 respectively, displaying the characteristic of the multi-valency formation of Se.[17] The bond length of 2.620 Å is smaller than that of α-tellurene, as the stronger interactions between Se atoms than Te. As usual, the smaller bond length in the same main Group, the larger $\kappa_L$. The phonon spectra are also explored and shown in Fig. 1(b), without any negative frequency, indicating the dynamical stability of the structure. The remarkable overlapping of acoustic and optical phonons can be found, implying strong optical-acoustic phonon scattering which will suppress the lattice thermal conductivity.[36]

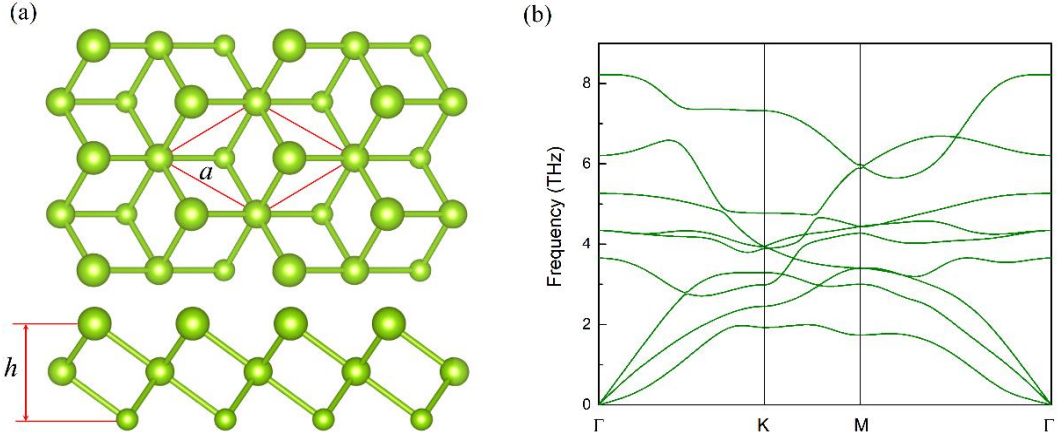

Fig. 1. The structure (a) and phonon dispersion (b) of 2D α-selenene. In (a), top view and side view are displayed, while the primitive cell is indicated by red solid lines in top view.

The group velocities around Γ point, Debye temperature $\theta_D$, and 2D elastic properties for α-selenene are calculated, as listed in Table 1. Here $\theta_D$ is obtained by $\theta_D = \hbar\omega_{max}/k_B$, where $\omega_{max}$ is the maximum of acoustic phonon frequency.[32,37] The 2D Young's modulus $E$ and Shear modulus $v$ are calculated based on elastic constants with the following expressions:[38]

$$E = (C_{11}^2 - C_{12}^2)/C_{11}$$
$$v = C_{12}/C_{11} \quad (2)$$

For comparison, these physical quantities of α-tellurene are also calculated and listed. The group velocities are quite small, comparable to those of 2D SnSe, arsenene and stanene,[32-36,39] but larger than α-tellurene.[17] The Debye temperature $\theta_D$ is larger than 2D SnSe, β-tellurene,[32] and α-tellurene. The 2D Young's modulus of α-selenene is larger than α-tellurene. Interestingly, the 2D Young's moduli of α-selenene and α-tellurene are quite closed to β-arsenene and β-antimonene respectively,[40] which are neighbors in the periodic table, indicating the similar in-plane stiffness. In the whole, the group velocities, elastic properties, and $\theta_D$ decrease from selenene to tellurene, as the interatomic interactions decrease. Usually, higher $\theta_D$ means higher thermal conductivity for materials with similar structure, as it indicates larger group velocity and frequency of phonons.[41] However, we will find there is an abnormal case for α-selenene and α-tellurene.

Table 1. Group velocity (in km/s), $\theta_D$, 2D Young's modulus and Shear modulus (in N m$^{-1}$) of α-selenene. For comparison, these properties of α-tellurene are also listed here.

|  | $v_{ZA}$ | $v_{TA}$ | $v_{LA}$ | $\theta_D$ | $E$ | $G$ |
|---|---|---|---|---|---|---|
| α-selenene | 0.80 | 2.4 | 3.5 | 163 | 55.1 | 21.9 |
| α-tellurene | 0.73 | 1.7 | 2.5 | 108 | 35.2 | 13.6 |

It should be noted that an effective thickness should be defined to calculate the thermal properties for 2D materials. Here, the thickness is defined as the summation of the buckling height $h$ and the twice of van der Waals radii of Se atoms.[32,42] Thus, we get the value of 6.913 Å for the effective thickness of α-selenene. With the thickness, the temperature dependent intrinsic lattice thermal conductivity $\kappa_L$ is calculated by the iterative scheme of ShengBTE,[30] exhibited in Fig. 2(a). The convergence test of $\kappa_L$ varying with the cutoff distance is in the inset of Fig. 1(a). It is found the $\kappa_L$ becomes convergent when the cutoff is 10$^{th}$ nearest neighbor. The intrinsic $\kappa_L$ shows inverse dependence of $T$, due to the stronger phonon–phonon scattering at higher temperature. We obtain the value of 3.04 W m$^{-1}$ K$^{-1}$ for the $\kappa_L$ of α-selenene at 300 K, even lower than the one of bulk Se,[43] but higher than square selenene.[15] For comparison, the $\kappa_L$ of α-tellurene from Ref. 17 is also displayed in Fig. 2(a). It is found α-selenene has much lower $\kappa_L$ than α-tellurene in the whole temperature range. For instance, the value of $\kappa_L$ is 9.85 W m$^{-1}$ K$^{-1}$ for α-tellurene at room temperature,[17] much higher than the one for α-selenene. This is quite different from our observation that α-selenene has a larger $\theta_D$ compared with α-tellurene. Based on the conventional theory, α-selenene should have higher $\kappa_L$ than α-tellurene, as usually higher $\theta_D$ means higher $\kappa_L$. The behind physical mechanics of this unusual behavior will be explained detailed latter.

To investigate the contributions of phonons with different frequencies to the total $\kappa_L$, the frequency-resolved $\kappa_L$ for α-selenene at 300 K is displayed in Fig. 2(b). The highest peak near about 1.8 THz indicates acoustic phonons have the most significant contribution to $\kappa_L$. The main of $\kappa_L$ comes from phonons lower than 3.5 THz, while nearly all of these phonons belong to acoustic branches as shown in Fig. 1(b). Thus, the acoustic phonons dominate the thermal transport of α-selenene. However, part of the contribution to $\kappa_L$ comes from optical phonons, as there is also a peak near 5.5 THz.

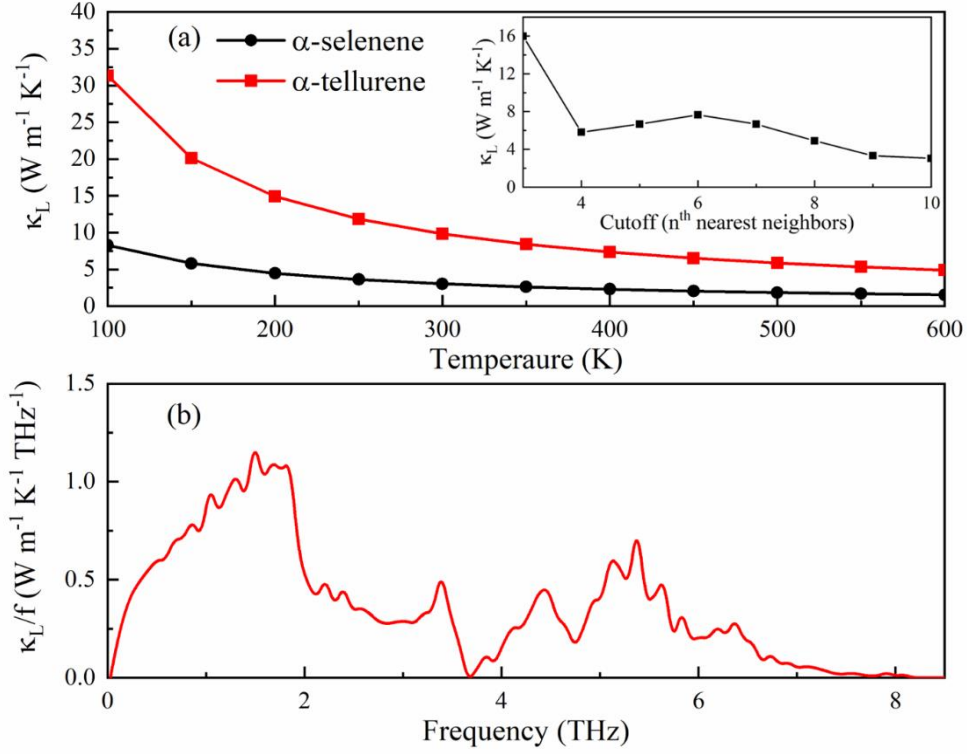

Fig. 2. The temperature dependent $\kappa_L$ (a) and frequency-resolved $\kappa_L$ for 2D α-selenene at 300 K (b). In (a) the $\kappa_L$ of α-tellurene from Ref 17 are also shown for comparison. And the inset of (a) shows the convergence of the $\kappa_L$ with cutoff distance.

In fact, all materials in practical applications have finite size, and the $\kappa_L$ will be suppressed by boundary scattering of different sample size significantly, especially at nanoscale. The boundary scattering for a phonon of branch $\lambda$ and wave vector **q** can be expressed as: $\frac{1}{\tau^B_{\lambda,q}} = \frac{v_{\lambda,q}}{L}$, where $L$ is the size of a material, $v_{\lambda,q}$ is group velocity. Note here the expression is corresponding to a completely diffusive boundary scattering of phonons. The size dependent normalized $\kappa_L$ at 300 K is calculated and exhibited in Fig. 3(a). The normalized $\kappa_L$ shows sensitive dependence on the variation of $L$ when $L$ is smaller than 0.1 $\mu$m, indicating the strong boundary scattering. For instance, with a sample size of 5 nm and 0.1 $\mu$m, $\kappa_L$ reaches about 50% and 94% of value for the infinitely large system, respectively. Whereas, it is nearly independent of $L$ above 0.5 $\mu$m. In fact, the normalized $\kappa_L$ reaches 0.99 with this size. This indicates that the remarkable size effect occurs when the size of α-selenene is less than 0.5 $\mu$m, much smaller than many other 2D materials such as 2D WTe$_2$ and α-tellurene.[17,44]

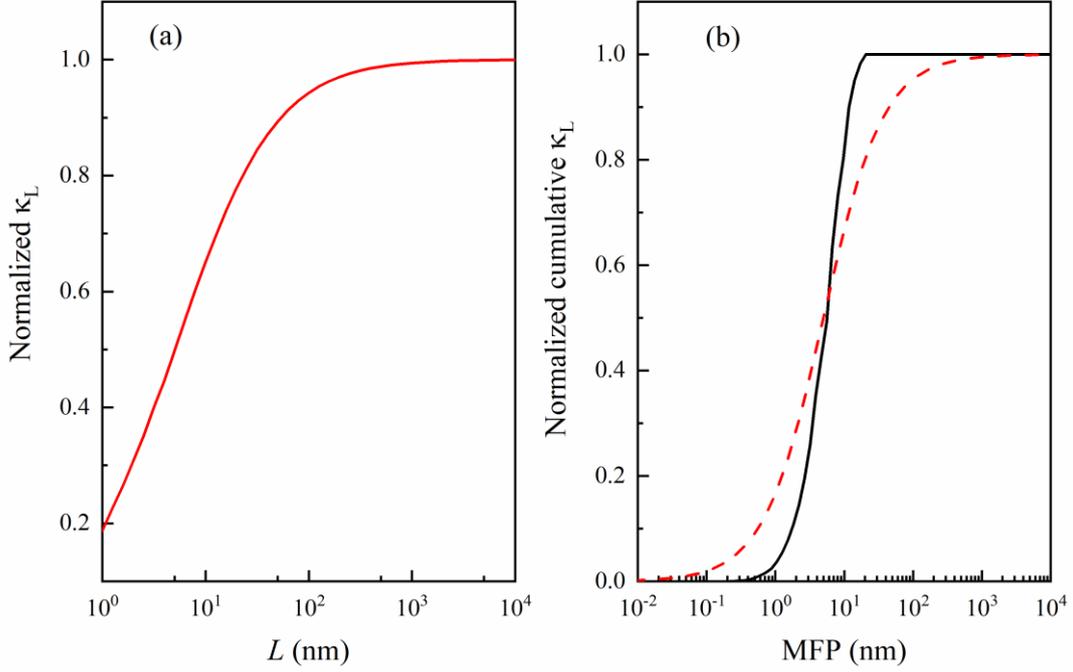

Fig. 3. Normalized 2D α-selenene $\kappa_L$ as a function of sample size $L$ at 300 K (a) and the MFPs dependent normalized cumulative $\kappa_L$ (b). The dashed red line in (b) represents the data of fitting.

Furthermore, the normalized cumulative $\kappa_L$ also can be used to estimate the size effect of materials, which is corresponding to a certain threshold of mean free path (MFP) $\Lambda_{max}$, and only the contributions of phonons with lower MFPs are taken into account. The normalized cumulative $\kappa_L$ at room temperature as a function of MPF is plotted in Fig. 3(b). The phonons with MFPs of 1 to 50 nm contribute mainly to the $\kappa_L$. Additionally, cumulative $\kappa_L$ can be fitted by a single parametric function: $\kappa_L(\Lambda \leq \Lambda_{max}) = \frac{\kappa_{max}}{1+\Lambda_0/\Lambda_{max}}$, where $\kappa_{max}$ means the intrinsic $\kappa_L$ with infinite size, and $\Lambda_0$ is a fitting parameter, which means the representative of the mean free path of relevant heat-carrying in materials.[30,44] Obviously, the fitting line reproduces the line of the calculated data reasonably well. We find the value of 5 nm for $\Lambda_0$, implying the $\kappa_L$ of α-selenene will decrease fast when the sample size smaller than several nanometers. The value is in reasonable agreement with Fig. 3(a), where the normalized $\kappa_L$ is about 0.5 for the material with the same size. This information is useful for reducing $\kappa_L$ in designing thermoelectric devices in the nanoscale.

Based on Eq. (1), group velocity $v_g$ and relaxation time $\tau$ of phonons are related to intrinsic $\kappa_L$ closely. These data at 300 K are calculated and shown in Fig. 4. The percentage contributions of ZA, TA, LA and optical branches to the total intrinsic $\kappa_L$ are 30.5, 17.7, 14.4 and 37.4%, respectively. After carefully checking and analyzing, we find the main factor which affects the thermal transport is $\tau$. In Fig. 4(a), it is found the group velocities of phonon branches don't have great differences, whereas the most contributions to total $\kappa_L$ are from three acoustic branches, and all the six optical modes contributes less than 40% to total $\kappa_L$. The domination of acoustic phonons comes from their high value of $\tau$ as shown in Fig. 4(b), especially those below 3.5 THz, which is also in agreement with Fig. 2(b). The $\tau$ of ZA mode is the highest of all the phonon mode, which is as high as about 20 ps. It determines ZA phonons contribute most to the total $\kappa_L$, though their group velocities are lower than other acoustic phonons. Compared with acoustic phonons, optical phonons don't contribute mainly to the total $\kappa_L$, since the $\tau$ of most optical phonons are in the order of $10^0$ ps only. Additionally, $\tau$ of a great number of phonons around 3 THz have a sharp drop, indicating the intense acoustic-optical phonon scattering near the frequency, as there is no gap between acoustic and optical branches. And the intrinsic $\kappa_L$ will be also suppressed by the phenomenon.[45] In the whole, the phonon group velocities in α-selenene are higher than those of α-tellurene, whereas the $\tau$ are at least an order smaller than α-tellurene especially for acoustic phonons with low frequency,[17] which results in lower intrinsic $\kappa_L$ in α-selenene than α-tellurene. It further confirms that $\tau$ of acoustic branches plays a key role in the thermal transport of a material.

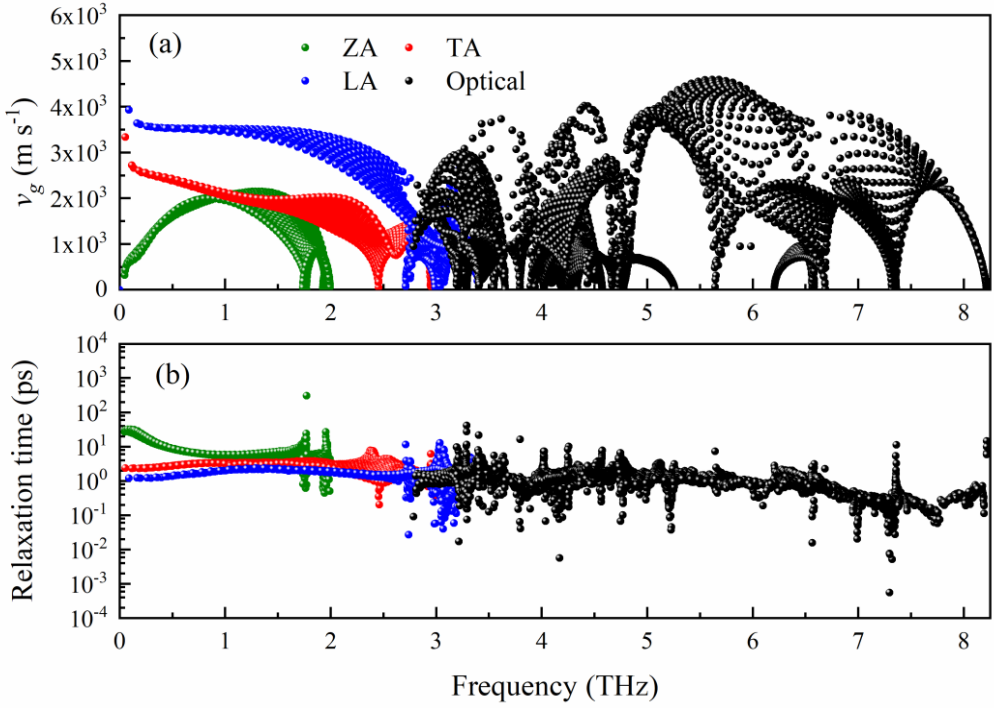

Fig. 4. Group velocity of phonons (a) and relaxation time (b) for 2D α-selenene. The green, red, blue and black markers represent ZA, TA, LA and optical phonons, respectively.

To further investigate the behind physical mechanics that leads to lower relaxation time τ as well as the lower intrinsic $κ_L$ in α-selenene, the total phase space for three-phonon processes $P_3$ and the Grüneisen parameter γ are calculated, as exhibited in Fig. 5(a) and (c). Note $P_3$ and γ of α-tellurene from Ref 17 are also plotted in Fig 5(b) and (d) for comparison. A limited $κ_L$ originates from the phonon-phonon scattering,[46] and $P_3$ can be directly used to estimate the number of three-phonon scattering processes available to each phonon, which must satisfy both energy and quasi-momentum conservations.[30,47,48] And it only depends on the phonon dispersions. Corresponding to each individual scattering channel, the inverse of τ, i.e., the phonon scattering rate, is proportional to $P_3$ and the square of corresponding scattering matrix element, and the latter reflects the phonon anharmonicity, which can be represented by γ. Thus, we can conclude large $P_3$ and γ lead to low $κ_L$. As in Fig. 5, $P_3$ in α-selenene are in the range of 5 to 35×$10^{-8}$, lower than those of α-tellurene, especially for the phonons with low frequency.[17] Obviously, it is not the factor which results in lower $κ_L$ in α-selenene. The Grüneisen parameter γ provides important information on thermal transport, as it can be used to measure the anharmonicity of phonons. γ for α-selenene are displayed in Fig. 5(c). In the whole, they are mainly in the range of -15 to 60, wider than the range of γ for α-tellurene.[17] γ of ZA phonons disperse widely for both α-selenene and α-tellurene. However, γ of LA and TA modes for α-selenene are much larger than those of α-tellurene. It indicates much stronger anharmonicity of LA and TA modes in α-selenene. As a matter of fact, thermal conductivity of a crystal structure is an outcome of summation of harmonic and anharmonic properties. Thus, it is reasonably concluded that the strong anharmonicity of phonons especially LA and TA phonons indeed results in the abnormal phenomenon of lower $\kappa_L$. Additionally, the γ near 3 THz have a sudden jump, where acoustic and optical phonons mix together. It indicates the intense acoustic-optical scattering, as there is no phonon gap between acoustic and optical branches. The phenomenon is in agreement with the dispersion of τ in Fig. 4(b), also representing suppressed intrinsic $\kappa_L$.[45]

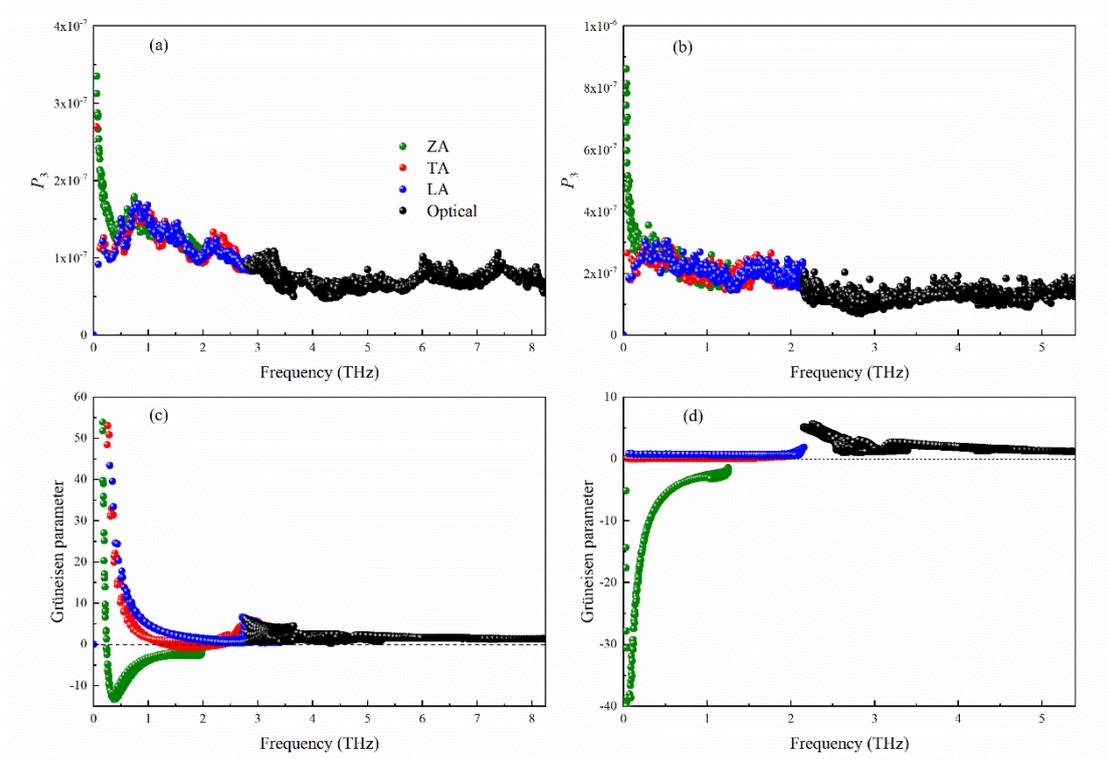

Fig. 5. $P_3$ (a) and Grüneisen parameters (c) for 2D α-selenene. For comparison, those for 2D α-tellurene are in (b) and (d).

## Conclusions

In summary, the intrinsic $\kappa_L$ of 2D α-selenene is investigated by using first-principles calculations theoretically. It is found α-selenene possesses an unusually low thermal conductivity of 3.04 W m$^{-1}$ K$^{-1}$, even lower than that of α-tellurene, which has similar structure and lower Debye temperature. This abnormal phenomenon mainly stems from the strong anharmonicity of LA and TA phonons, as shown by Grüneisen parameters. Its phonon spectra, mechanical properties, Debye temperature are also exhibited. The size effect and boundary effect are studied by cumulative $\kappa_L$ and size-dependent $\kappa_L$ respectively. Furthermore, the total phase space of three-phonon processes $P_3$, the group velocities and relaxation times of phonons are investigated to explore the behind mechanism of low thermal conductivity. The relatively low relaxation times mainly determine the low thermal conductivity. It is concluded reasonably that relaxation time plays the dominant role for the thermal transport of α-selenene.

## Author information

Corresponding Authors
*Email: liugang8105@gmail.com
* Email: zhibin.gao@nus.edu.sg
ORCID
Gang Liu: 0000-0001-9033-7376
Zhibin Gao: 0000-0002-6843-381X

## Acknowledgments

This work is supported by the National Natural Science Foundation of China (Nos. 11974100, 61764001 and U1404212). Z. Gao acknowledges the financial support from MOE tier 1 funding of NUS Faculty of Science, Singapore (Grant No. R-144-000-402-114).